# Revisiting the foundations of the quantum theory of atoms in molecules: Some open problems

Shant Shahbazian

*Department of Physics, Shahid Beheshti University, G. C., Evin, Tehran, Iran, 19839, P.O. Box 19395-4716.*

Tel/Fax: 98-21-22431661

E-mail: sh_shahbazian@sbu.ac.ir

## Abstract

In a series of papers in the last ten years, various aspects of the mathematical foundations of the quantum theory of atoms in molecules have been considered by this author and his coworkers in some details. Although these considerations answered part of the questions raised by some critics on the mathematical foundations of the quantum theory of atoms in molecules, however, new mathematical problems also emerged during these studies that were reviewed elsewhere [Int. J. Quantum Chem. **111**, 4497 (2011)]. Beyond mathematical subtleties of the formalism that were the original motivation for initial exchanges and disputes, the questions raised by critics had a constructive effect and prompted the author to propose a novel extension of the theory, now called the multi-component quantum theory of atoms in molecules [Theor. Chem. Acc. **132**, 1365 (2013)]. Taking this background into account, in this paper a new set of open problems is put forward that the author believes proper answers to these questions, may open new doors for future theoretical developments of the quantum theory of atoms in molecules. Accordingly, rather than emphasizing on the rigorous mathematical formulation, the practical motivations behind proposing these questions are discussed in detail and the relevant literature are reviewed while when possible, evidence and routes to answers are also provided. The author hopes that proposing these open questions as a compact package may motivate more mathematically oriented people to participate in future developments of the quantum theory of atoms in molecules and its multi-component version.

## Keywords

# 1. Introduction

Since the publication of the seminal monograph of Bader on the quantum theory of atoms in molecules (QTAIM) [1], which summarizes the works of Bader and his research group before 1990, the QTAIM with all its pros and cons has been widely conceived as the comprehensive theory of atoms in molecules. To make the theory more accessible to general audience, Popelier's book that appeared in 2000 tries to act as a textbook for newcomers to this field, though, occasionally, one may also find Popelier's own interesting insights in it as well [2]. More recently, in 2007, the editorial compendium of Matta and Boyd tries to cover, beyond doubt successfully, a wide range of the QTAIM's applications from solid state physics to chemistry and then to molecular biology in a coherent way [3]. All in all, the QTAIM is now part of the toolbox that aims to tackle various problems where the AIM structure is of prime importance [4,5]. However, in contrast to its widespread recognition, sporadically, the mathematical foundations as well as interpretation of the QTAIM analysis both had been a matter of disputes and exchanges [6-17]; in this study, only the former is in focus while for interpretive problems, the interested reader may consult [18,19] and references cited therein. Starting from a decade ago, inspired by these exchanges, the present author and his coworkers reconsidered the mathematical foundations of the QTAIM in detail trying to resolve some issues put forward by critics of the QTAIM formalism [20-29]. Interestingly, as a byproduct, this reconsideration eventually gave rise to an extended version of the QTAIM, finally termed the multi-component QTAIM, which widens the applications of the AIM analysis beyond the Born-Oppenheimer paradigm as well as exotic molecular species [30-45].

Although some of the original problems related to the QTAIM formalism, which have been discussed elsewhere [6,7,46], seem to be addressed properly in our reconsiderations, some still remain problematic and needs further scrutinizing. On the other hand, one may propose new problems that seem to be unnoticed (or at least not stated officially) and hopefully, will open new doors for future developments of the QTAIM. Accordingly, the paper is organized based on a handful of selected open questions and each subsection in next section discusses one problem and the relevant background and literature. The author has proposed some leads where it is possible, which may eventually help to solve the corresponding problem or at least to be used as a starting point for future considerations. Also, let us stress that the style of presentation of problems is intentionally to some extent informal and far from mathematically rigorous treatment making the whole subject accessible to a larger range of audience. It is important to stress that the motivation behind presenting these problems is their potential for new developments rather than filling holes in previous chains of mathematical reasoning of the conventional formalism. The author hopes that this paper may encourage more theoreticians from various fields to participate in future developments of the QTAIM and its extended multi-component version.

## 2. Open problems and corresponding background

### 2.1. What is the very origin of the topological atom?

The very notion of the atomic basin, usually termed a "topological atom" [1], is fundamental to the QTAIM. One may quote directly from Bader's monograph: "Atoms exist in real space and they are defined by a partitioning of real space" (see page 55 in [1]) and: "An atom, free or bound, is defined as the union of an attractor and its associated basin" (see page 28 in [1]).  The latter is the most basic definition of an atom

(or AIM) within the context of the QTAIM that employs the gradient vector field of the one-electron density, $\vec{\nabla}\rho(\vec{r})$, where $\rho(\vec{r}) = N \int d\tau' \Psi^*\Psi$ ($N$ is the number of the

electrons and $\Psi$ is the wavefunction of the system while $d\tau'$ implies summing over

spin variables of all electrons and integrating over spatial coordinates of all electrons

except one of them).  The vector field reveals maxima, termed also (3, -3) critical points

[1], at the position of nuclei, which act as the attractors of the gradient field, while all the

gradient paths terminating at an attractor disclose the basin of the attractor, i.e. the atomic basin.  The resulting partitioning of the real (3D) space is exhaustive without any overlap between the AIM and the boundaries between AIM, i.e. the interatomic-surfaces, are delineated by the zero-flux surfaces containing all the gradient paths that terminate at (3,

-1) critical points.  These critical points are saddle points which are maximum at two directions and minimum along the other [1].  The interatomic-surfaces have the basic property that $\vec{\nabla}\rho(\vec{r})$ is perpendicular to the unit vector normal to the surface, i.e.

$\vec{\nabla}\rho(\vec{r}).\vec{n}(\vec{r}) = 0$; this equation, termed the local zero-flux equation, is usually

introduced as the basic equation governing the interatomic-surfaces [1,47].  Figure 1

depicts all these ingredients schematically in a typical diatomic molecule as an illustrative example.

      The first four chapters of Bader's monograph thoroughly consider the topology of

the one-electron density, i.e. analyzing $\vec{\nabla}\rho(\vec{r})$, as well as the topological definition of

AIM.  In chapters 5, 6 and 8 the author goes in length to demonstrate that this definition

fulfills the requirements to be also the "quantum atom", based on the so-called *subsystem*

*quantum mechanics* developed within the context of the QTAIM, which tries to extend many theorems of quantum mechanics to *real-space quantum subsystems* [1]. Accordingly, a detailed reading of the monograph reveals three independent "derivations" of the local zero-flux equation within the context of the subsystem quantum mechanics which are: 1- Through the uniqueness of the kinetic energy of the AIM (see page 148 [1] while for a critical attitude toward the uniqueness claim see [46]), 2- Through the subsystem regional virial theorem (see pages 176 and 177 in [1]), 3- Through the subsystem variational procedure (see pages 158 and 378 in [1]). All these supposed derivations finally yield the following equation: $\oint_{\partial\Omega} dS \ \vec{\nabla}\rho(\vec{r}).\vec{n}(\vec{r}) = 0$, which is termed the net zero-flux equation; through Gauss's theorem, the net zero-flux equation may be alternatively written as: $\int_{\Omega} d\vec{r} \ \nabla^2\rho(\vec{r}) = 0$. The local zero-flux implies the net zero-flux but the reverse is *not* true, which was known to Bader as early as 1971 [48,49], and since then has been "rediscovered" subsequently many times [12,50-52]. More recently, we have established computationally that certain 3D "objects", collectively termed *quantum divided basins*, are derivable from the net zero-flux equation, which from "morphological" viewpoint can be similar or quite distinct from the topological atoms [22,24,26]. The presence of the quantum divided basins reveals that the topological atoms are *not uniquely* derivable from the net zero-flux equation and they are just a tiny subset of the whole 3D objects satisfying the net zero-flux equation. In other words, the objects compatible to the mathematical fabric of the subsystem quantum mechanics are not exclusively the topological atoms. Thus, the question arises that what makes the topological atoms distinct?

One answer to this question is that they are the *natural* objects for an AIM theory since there are simplest objects resembling most to what one expects intuitionally from an AIM while in general the quantum divided basins do not meet this criterion of "naturalness". In other words, each topological atom by its very definition contains a nucleus and part of surrounding one-electron density with a small imbalance of the positive nuclear and the negative electronic charges making it a semi-neutral object; this picture fits to the general conception of the AIM, which is a marginally "modified" version of a free atom [53]. In this viewpoint, the concept of the topological atom must be assumed as an independent "axiom" of the QTAIM formalism in addition to the axioms used to introduce the subsystem quantum mechanics. For practical purposes, the naturalness seems to be a satisfying answer however, for one interested to link the concepts of the QTAIM to quantum mechanics this naturalness though seems reasonable phenomenologically, has a more or less *ad hoc* character. A qualitative answer to the origin of the naturalness has been provided by Bader himself: "… Dominant among these is the attractive force exerted by the nuclei, a consequence of the localized nature of the nuclear charge. This interaction is responsible for the single most important topological property exhibited by a molecular charge distribution of a many-electron system-that, in general, $\rho(\vec{r}; X)$ exhibits local maxima only at the positions of the nuclei." (see page 14 in [1]). This reasoning was implicitly confirmed when we tried to apply the QTAIM analysis to a hypothetical model of molecule where the coulombic interactions between electrons and nuclei were replaced by the harmonic potential, i.e. the Hookean model of molecule [28]. The results of the analysis on the Hookean $H_2^+$ demonstrated that instead of two attractors in $\vec{\nabla}\rho(\vec{r})$ on each of the nuclei, a single attractor emerges in the middle of the nuclei [28]. Evidently, the fact that the attractors of $\vec{\nabla}\rho\left(\vec{r}\right)$ in the coulombic systems are usually revealed only on the nuclei is not something trivial and is tied intrinsically to the nature of the interaction of nuclei and electrons. Accordingly, the naturalness problem may be cast in a more concrete language: How the coulombic interactions between nuclei and electrons dictate the topology of the one-electron density in a way that the concept of the topological atom arises naturally?

To answer this question, it seems legitimate to consider the one-electron density of free atoms at first step. It is well-known computationally that the ground state $\vec{\nabla}\rho_{atom}\left(\vec{r}\right)$ of a free atom reveals a single attractor, a (3, -3) critical point, at the position of nucleus while $\rho_{atom}\left(\vec{r}\right)$ decays "monotonically" far from the nucleus [1,54,55] (check Figure 1), however, from theoretical viewpoint, this is not a proven conjecture yet [56-59]. On the other hand, if the free atom's one-electron density is perturbed marginally in a molecule because of interaction with other AIM, it seems reasonable to assume that the one-electron density of a molecule is nearly the sum of the one-electron densities of the constituent free atoms, $\rho_{mol}\left(\vec{r}\right) \approx \sum_{i}\rho_{atom,i}\left(\vec{r}\right)$, which is officially called the promolecule model of the one-electron density (for examples as well as a recent bibliography on this model see [60,61]). At first glance it seems that within the context of the promolecule model a molecule containing $q$ nuclei, also contains $q$ number of (3, -3) critical points at the position of the clamped nuclei. However, many computational as well as some experimental studies demonstrated that (3, -3) critical points may also appear sporadically at positions far from nuclei usually in the middle of two nuclei [1,62-72]. There have been proposals to get rid of these (3, -3) critical points, sometimes called non-nuclear attractors, and the corresponding "pseudo-atoms" [73]. However, the elegant study by Pendas and coworkers demonstrated that these pseudo-atoms arise in general when the distance between two nuclei is smaller than a critical value and are not an artifact of computational procedures [74]. Usually, these critical distances are smaller than the equilibrium distances thus, the pseudo-atoms seem deceptively sporadic but upon applying hydrostatic pressures and corresponding contraction of the inter-nuclear distances, pseudo-atoms appear in the AIM structure of every molecular system [75]. One may conclude that the naturalness of the topological atoms stems from the fact that in general, for each pair of nuclei the critical inter-nuclear distance is smaller than the inter-nuclear distance at the equilibrium geometry. Based on this background, let us now formulate the idea of the naturalness of the topological atoms as a mathematical conjecture:

**Conjecture**: *For a molecular system containing $N$ electrons (with position vectors $\{\vec{r}\}$) and $q$ clamped nuclei (with position vectors $\{\vec{R}\}$ and inter-nuclear distances $R_{\alpha\beta} = \left|\vec{R}_\alpha - \vec{R}_\beta\right|$) described by the following electronic Hamiltonian:*

$$\hat{H} = \left(\frac{-\hbar^2}{2m_e}\right)\sum_i^N \nabla_i^2 + \sum_i^N \sum_{j>i}^N \frac{1}{r_{ij}} + \hat{V}_{ext} \ , \qquad \hat{V}_{ext} = -\sum_i^N \sum_\alpha^q \frac{Z_\alpha}{\left|\vec{R}_\alpha - \vec{r}_i\right|},$$

*There is always a critical distance between each pair of nuclei, denoted as $\left\{R_{\alpha\beta}^c\right\}$, that for geometries for which the inter-nuclear distances are larger than $\left\{R_{\alpha\beta}^c\right\}$,*

*$\vec{\nabla}\rho_{mol}\left(\vec{r};\left\{R_{\alpha\beta} > R_{\alpha\beta}^c\right\}\right)$ contains just $q$ number of (3, -3) critical points at the position* *of the nuclei while for "most" molecules:* $\left\{ R_{\alpha\beta}^{eq} > R_{\alpha\beta}^{c} \right\}$, *where* $\left\{ R_{\alpha\beta}^{eq} \right\}$ *is the set of the inter-nuclear distances at the equilibrium geometry.*

Some points regarding this conjecture are worth emphasizing. As discussed, this conjecture is closely related to the unproven conjecture of the monotonicity of $\rho_{atom}\left( \vec{r} \right)$ and probably one needs to prove the monotonicity before proving the present conjecture. Also, it is interesting to extend the conjecture to a larger class of external potentials, e.g. those derived from the finite nucleus models [76], since the topology of the one-electron density is probably independent from marginal modifications of the external potential (for a similar proposal see [23]). Evidently, the last part of the conjecture, namely the fact that in "most" molecules the equilibrium inter-nuclear distances are larger than the critical distances, is hard to properly paraphrased mathematically and probably hardest part to be proved. Let us finally emphasize that this conjecture tacitly implies that the "efficiency" of the concept of the real-space quantum subsystem, e.g. the concept of the topological atom, is intrinsically tied to the nature of interactions of the constituents of the molecular system and cannot be extended *per se* to other many-body quantum systems (for possible extensions to non-coulombic systems see [28,29]).

## 2.2. How to extend the concept of topological atom's "energy" for non-coulombic interactions?

It has been demonstrated previously in detail that most of the formalism and theorems of the subsystem quantum mechanics are independent from the nature of the interactions of the constituents of the quantum system [23,28,29]. In other words, the nature of the potential energy function of the Hamiltonian describing the physics of system is not relevant to the abstract results derived within the context of the subsystem quantum mechanics. Two important pillars of the subsystem quantum mechanics namely, the subsystem variational procedure [23], and the subsystem hypervirial theorem [28], are concrete examples in this regard. However, as emphasized recently [29], the original procedure of introducing the energy of a topological atom through using the subsystem virial theorem explicitly employs the *homogeneity* of the coulombic potential energy function (see pages 186-191 in [1]). Any marginal modification of the coulombic potential energy in a molecular system, e.g. using any of the finite nucleus models instead of the point charge model [76] or adding terms corresponding to various small magnetic interactions [77,78], destroys the homogeneity of the potential energy function, making the originally proposed procedure useless. Accordingly, the whole procedure of introducing topological atom's energy will be untenable in such cases even if absolute deviations from the coulombic potential to be small. Although for most (but not all) practical purposes one may neglect deviations from the coulombic interactions in a molecular system, for one believing that the AIM are "real" entities [79], confining any ingredient of the QTAIM just to the coulombic potential is something artificial. Accordingly, it seems reasonable that for the AIM, to be a "generic" property of molecule [1], the QTAIM formalism must be able to cope with diverse families of "effective" molecular Hamiltonians that the coulombic Hamiltonian is just an example. Though we have recently extended the procedure of introducing energy for real-space subsystems to the whole family of the homogeneous potential energy functions [29], as will be discussed, this procedure is not extendable to inhomogeneous potential energy functions. Let us consider the extended procedure briefly in first place and then the obstacles emerging from its extension to inhomogeneous potential energy functions.

In order to start the analysis, imagine an $N$-particle quantum system composed of indistinguishable particles that upon solving corresponding Schrödinger's equation, $\hat{H}\Psi = E\Psi$, at least one bound state emerges and is at the mechanical equilibrium, i.e. the Hellmann-Feynman (HF) forces are null [1]. The fact that the HF forces are assumed to be null from the outset of the analysis restricts the whole analysis only to the equilibrium geometries. This is somehow annoying since because of nuclear vibrations as well as chemical reactions the QTAIM analysis of non-equilibrium geometries is also desirable. There have been some attempts to extend the analysis beyond equilibrium geometries when the HF forces are not null and these have been summarized by Keith (see chapter 3 in [3]). But, none of the proposed schemes is satisfying as stressed by the author himself (a possible tentative route to solve this problem have been proposed by present author recently that needs to be worked out in detail [45]). Throughout present analysis we will restrict our analysis only to the equilibrium geometries and do not try to consider the more involved problem of generalizing the analysis to non-equilibrium geometries. Let us also assume that the potential energy function, $V(\vec{r}_1, \vec{r}_2, ..., \vec{r}_N)$, has the following property: $V(s\vec{r}_1, s\vec{r}_2, ..., s\vec{r}_N) = s^n V(\vec{r}_1, \vec{r}_2, ..., \vec{r}_N)$ where $\vec{r}_i$ is the position vector for $i-th$ particle and $s$ is an arbitrary scaling parameter while $n$ is called the degree of homogeneity. For this family of potential energy functions when the potential depends on the interparticle distances, i.e. $\left| \vec{r}_i - \vec{r}_j \right|$, delineating the contribution of each quantum particle from the total potential energy is non-trivial [1]. The general strategy is introducing a projection operator, $\hat{P}_i$, with the following property: $\sum_i \hat{P}_i V = V$, where $\hat{P}_i V$ is the contribution of the $i-th$ particle. In the case of above introduced family of potential energy functions the projection operator is as follows: $\hat{P}_i = \vec{r}_i \cdot \vec{\nabla}_i / n$, where the coulombic potential is a special case: $n = -1$. The Hamiltonian of the system is written based on the projection operators as follows:

$$\hat{H} = \hat{T} + \hat{V} = \sum_i^N \hat{t}_i + (1/n)\vec{r}_i \cdot \vec{\nabla}_i V = \sum_i^N \hat{h}_i \text{, where } \hat{t}_i = \left(-\hbar^2/2m\right)\nabla_i^2 \text{ (}m\text{ is the mass}$$

of particles) is the kinetic operator of the $i-th$ particle and $\hat{h}_i = \hat{t}_i + (1/n)\vec{r}_i \cdot \vec{\nabla}_i V$. Using the fact that the Hamiltonian is now expressed based on the one-particle contributions, the energy density of the system is introduced as follows:

$$E\left(\vec{r}_1\right) = N \int d\tau' \, \Psi^* \hat{h}_1 \Psi + e\left(\vec{r}_1\right) \text{ (we will drop subscript 1 afterwards from all}$$

equations since the particles are assumed to be indistinguishable). The second term in the right-hand side of the equation, $e\left(\vec{r}\right)$, has been added with the property: $\int d\vec{r} \, e\left(\vec{r}\right) = 0$, and is undetermined as far as one imposes only: $\int d\vec{r} \, E\left(\vec{r}\right) = E$, where $E$ is the total energy of the system. This is a way to stress that in contrast to the total energy of a molecule, which is uniquely defined expect from an additive constant, the energy density is not uniquely defined as stressed and discussed by Ayers and coworkers in their comprehensive study on the kinetic energy densities [46]. The energy density may be decomposed further into the kinetic energy density, $K\left(\vec{r}\right) = N \int d\tau' \, \Psi^* \hat{t} \Psi$, and the virial energy density, $V^B\left(\vec{r}\right) = N \int d\tau' \, \Psi^* \left(-\vec{r} \cdot \vec{\nabla} V\right) \Psi$, thus:

$$E\left(\vec{r}\right) = K\left(\vec{r}\right) - (1/n)V^B\left(\vec{r}\right) + e\left(\vec{r}\right). \text{ In order to use the energy density to derive the}$$

energy of a real-space subsystem, denoted by spatial basin $\Omega$, $e(\vec{r})$ must be determined in a proper way and generally (in contrast to the total space): $\int_{\Omega} d\vec{q}\; e(\vec{r}) \neq 0$. At this stage of development, the local version of the subsystem virial theorem must be used to determine $e(\vec{r})$, which is as follows:

$$2K(\vec{r}) = -V^B(\vec{r}) - V^S(\vec{r}) - \left(\hbar^2/4m\right)\nabla^2\rho(\vec{r}),$$ where $V^S(\vec{r}) = \vec{\nabla}.\left(\vec{r}\otimes\vec{\vec{\sigma}}(\vec{r})\right)$ is

called the surface virial density [1]. The integral of $V^S(\vec{r})$ on the whole space, i.e. $R^3$, is null: $V^S_{R^3} = \oint dS\left(\vec{r}\otimes\vec{\vec{\sigma}}(\vec{r})\right).\vec{n} = 0$, making it unimportant for the total system but its integral is generally non-zero for a real-space subsystem,

$V^S_{\Omega} = \oint_{\partial\Omega} dS\left(\vec{r}\otimes\vec{\vec{\sigma}}(\vec{r})\right).\vec{n} \neq 0$. Also, $\vec{\vec{\sigma}}(\vec{r})$ is the stress tensor density,

$$\vec{\vec{\sigma}}(\vec{r}) = \left(\frac{N\hbar^2}{4m}\right)\int d\tau' \left\{ \Psi^*\left(\vec{\nabla}\otimes\vec{\nabla}\Psi\right) + \Psi\left(\vec{\nabla}\otimes\vec{\nabla}\Psi^*\right) - \left(\vec{\nabla}\Psi^*\right)\otimes\left(\vec{\nabla}\Psi\right) - \left(\vec{\nabla}\Psi\right)\otimes\left(\vec{\nabla}\Psi^*\right) \right\}$$

while the symbol $\otimes$ has been used in the above equations to emphasize on the dyadic nature of the products [1]. It is timely to emphasize that the introduced stress tensor density is one among an infinite set of possible stress tensor densities and this is the deeper route toward the non-uniqueness of the energy density as discussed elsewhere [46]. For real-space quantum subsystems, the presence of the surface virial density guarantees that the total virial density, $V^B(\vec{r}) + V^S(\vec{r})$, is origin independent [1,28], which is a condition to be satisfied for any well-defined energy of a real-space subsystem. Accordingly, $e(\vec{r}) = -(1/n)V^S(\vec{r})$ is used hereafter and based on the subsystem virial theorem the energy density simplifies to:

$$E(\vec{r}) = (1 + 2/n) K(\vec{r}) + (\hbar^2/4mn) \nabla^2 \rho(\vec{r})$$ or alternatively to:

$$E(\vec{r}) = -(1/2 + 1/n)\{V^B(\vec{r}) + V^S(\vec{r})\} - (\hbar^2/8m) \nabla^2 \rho(\vec{r}).$$ The case of $n = -2$ at first glance may seem problematic but it is well-known that in the case of attractive interactions, corresponding total system does not have well-behaved bound states (see subsection 35 in [80]). The energy of a real-space quantum subsystem is now easily derivable from the integration of the energy density within the atomic basin, $E_\Omega = \int_\Omega d\vec{q} \ E(\vec{r})$, and the net zero-flux equation is assumed to be satisfied for any real-space subsystem, then: $E_\Omega = (1 + 2/n) K_\Omega = -(1/2 + 1/n)\{V^B_\Omega + V^S_\Omega\}$ (for applications of the formalism to non-coulombic systems see [29,45]). Finally, topological atom's energy in the coulombic systems is recovered as a special: $E_\Omega = -K_\Omega = (1/2)\{V^B_\Omega + V^S_\Omega\}$.

Let us now widen the family of considered potential energy functions assuming that $V(\vec{r}_1, \vec{r}_2, ..., \vec{r}_N)$ is not a homogeneous potential energy function but is composed of two functions, $V = V_1 + V_2$, with two different degrees of homogeneity, $n_1$ and $n_2$ [81]; an example relevant to our study is the case where one part is composed of the usual electric coulombic interactions and the other from the magnetic interactions between electrons [77,78]. The projection operators are introduced as follows: $V = \sum_i \hat{P}_i V_1 + \sum_i \hat{Q}_i V_2$, where $\hat{P}_i = \vec{r}_i \cdot \vec{\nabla}_i / n_1$ and $\hat{Q}_i = \vec{r}_i \cdot \vec{\nabla}_i / n_2$, and the Hamiltonian is written accordingly: $\hat{H} = \sum_i^N \hat{h}_i$, where $\hat{h}_i = \hat{t}_i + (1/n_1) \vec{r}_i \cdot \vec{\nabla}_i V_1 + (1/n_2) \vec{r}_i \cdot \vec{\nabla}_i V_2$.

The energy density is introduced with the same procedure used for the homogeneous functions: $E(\vec{r}) = K(\vec{r}) - (1/n_1)V_1^B(\vec{r}) - (1/n_2)V_2^B(\vec{r}) + e(\vec{r})$, where

$V_j^B(\vec{r}) = N \int d\tau' \, \Psi^*(-\vec{r} \cdot \vec{\nabla} V_j)\Psi$ ($j = 1, 2$). The local version of the subsystem

virial theorem is now: $2K(\vec{r}) = -V_1^B(\vec{r}) - V_2^B(\vec{r}) - V^S(\vec{r}) - (\hbar^2/4m)\nabla^2\rho(\vec{r})$ and

evidently, this equation cannot be used to propose a proper form for $e(\vec{r})$ to guarantee the origin independence of the topological atom's energy. Thus, the original procedure used to introduce topological atom's energy does not seem to be applicable even when small deviations from the coulombic interactions are conceived, which is clearly an unpleasant situation. The whole argument can be extended to potential energy functions separable to more than two homogenous functions while for more intricate potential energy functions, which are not separable to a finite number of homogeneous functions, the obstacle is worse. One immediate but *ad hoc* solution to this problem is assuming that the "balance" between the total kinetic, $T_{R^3}$, and the total potential energy, $V_{R^3}$, of a system, $V_{R^3}/T_{R^3} = \alpha$ and $E = (1 + \alpha)T_{R^3}$ ($\alpha$ must be derived from ab initio calculations) is also retained "locally": $E(\vec{r}) = (1 + \alpha)K(\vec{r})$. Obviously, in the case of the coulombic interactions, i.e. $\alpha = -2$, this equation yields the well-known result: $E_\Omega = -K_\Omega$, at the equilibrium geometries whereas in most real molecules deviations from $\alpha = -2$ is minute and this may probably justify the procedure. However, this procedure is too crude to be seen as a proper solution and we leave the proper definition of topological atom's energy upon marginal modification of the coulombic potential as an open problem.

## 2.3. What is the most efficient way of introducing the concept of the topological atom's "state"?

The concept of "state" is central in physics and by definition, knowing the state of a system means that all accessible information for the system is at hand, thus, it is legitimate to ask that how the state of a topological atom must be defined. The simplest answer is to list all properties, $\{M_\Omega\}$, e.g. the energy of a topological atom, or property densities, $\{M(\vec{r})\}$, e.g. the energy density of a topological atom, as the state of the atom. Then the question emerges whether these properties or property densities are really independent or they are somehow inter-dependent; simple examples confirming the latter view are the electron population and the atomic dipole moments, which are both derivable from the one-electron density of a topological atom (see pages 182 and 183 in [1]). Also, more intricate examples are the Ehrenfest forces operative on the atom and atom's energy, which are both derivable from the stress tensor density of the atom (see pages 174, 177 and 189 in [1]). As an alternative approach, inspired by the Hohenberg-Kohn (HK) theorem, which is the foundation of the density functional theory [82,83], Bader and Becker used the extension of the HK theorem to subspaces, proposed originally by Riess and Münch [84], to demonstrate that knowing the one-electron density of a topological atom, $\rho_\Omega(\vec{r})$, *in principle* determines uniquely all properties of the atom [85]. In this sense, it may be claimed that $\rho_\Omega(\vec{r})$ is *the* state of the topological atom, however, this is to some extent deceptive as will be discussed in the remaining of this subsection. Let us first reconsider the idea of Bader and Becker using the extended version of the Riess-Münch theorem namely, the holographic electron density (HED) theorem [86].

The HED theorem states that knowing the ground state one-electron density of a molecule in any arbitrary subspace, $\omega$, namely, $\rho_\omega(\vec{r})$, uniquely determines the one-electron density in the total space. One the other hand, roughly speaking, the HK theorem states that the one-electron density of the ground state uniquely determines all properties of the molecule thus the chain of reasoning goes as follows: $\rho_\omega(\vec{r}) \xrightarrow{HED} \rho(\vec{r}) \xrightarrow{HK} \hat{H}, \Psi$. Since there is no restriction on $\omega$ one may choose the subspace as the space occupied by the topological atom, $\omega = \Omega$, and then use the fact that knowing the molecular wavefunction suffices to determine the properties of the topological atoms: $\rho_\Omega(\vec{r}) \xrightarrow{HED} \rho(\vec{r}) \xrightarrow{HK} \hat{H}, \Psi \xrightarrow{QTAIM} M_\Omega$. To use the language of the density functional theory one may state that $M_\Omega$ is a functional of $\rho_\Omega(\vec{r})$ namely, $M_\Omega\left[\rho_\Omega(\vec{r})\right]$, and in this sense $\rho_\Omega(\vec{r})$ is indeed the state of the topological atom. Although this reasoning is in first glance convincing, according to the best of author's knowledge, the explicit form of the functionality has never been worked out so the proposed link has a formal nature with no real practical applications. In the density functional theory another theorem by Hohenberg and Kohn circumvents this problem by demonstrating that the variational principle of quantum mechanics may be rewritten employing $\rho(\vec{r})$ instead of $\Psi$ [82,83]. This opens the door for replacing Schrödinger's equation with the Kohn-Sham equations [87], both are in principle equivalent [83], making the computational implementation of density functional theory feasible. In contrast to an initial attempt [88], the idea of proposing a variational principle for the topological atoms, using $\rho_\Omega(\vec{r})$ or any other function, did not reach to somewhere [25], thus knowing $\rho_\Omega(\vec{r})$ does not let the deduction of $M_\Omega$. One other subtle problem with the above reasoning is the fact that imposing the restriction $\omega = \Omega$ is artificial and any volume in the interior of a topological atom or shared between topological atoms are equally legitimate to be used. In other words, by a small modification of the above reasoning, a $\omega$ within interior of a topological atom may formally determine the properties of another topological atom without any need of knowing something about $\rho_\Omega(\vec{r})$ of the target atom, which is clearly an unpleasant situation. Taking all these problems while the Bader and Beker's reasoning must be seen a step forward the question remains whether one may define the state of topological atom in a way that the properties of the atom may be deduced from this definition in practice. Taking the large and diverse literature on the open quantum systems, flourished in the last decades [89,90], the time is ripe to reconsider the problem of efficient definition of topological atom's state in the light of these new developments.

### 2.4. Do the interatomic-surfaces act as "hologram" for the topological atoms?

In recent years there has been a growing interest in "holographic principles" in certain disciplines of physics though they are usually far from the domain of atoms and molecules [91-93]. Such idea may have deep implications in fundamental physics revealing non-trivial links between seemingly unrelated theories [93]. Roughly speaking, a holographic principle makes a link between a theory formulated in $D$ spatial dimension with another theory formulated in $D+1$ dimension. Generally, at first glance the two theories may have nothing in common but the holographic principle enables one to "translate" concepts and calculations in one theory to concepts and calculations in the other theory. Interestingly, the mathematical structure of the QTAIM is naturally prone to a holographic viewpoint through the balance dictated by the subsystem hypervirial theorem between certain properties of the topological atoms and the corresponding properties of the interatomic-surfaces as will be discussed. In this context, it is possible to demonstrate that to determine many properties of the topological atoms just knowing what goes on the interatomic-surfaces suffices. Let us first consider a simple example using Gauss's theorem and then the subsystem hypervirial theorem as favorable examples and discussing the obstacles to formulate a general holographic principle within the context of the QTAIM.

The usual procedure to attribute properties to the topological atoms, i.e. their share from the observables of total system, is deriving a property density for each observable and then integrating these densities within the basins of the topological atoms delineated by the inter-atomic surfaces, $\int_{\Omega} d\vec{r} M(\vec{r}) = M_{\Omega}$ [1]. This integration means that the property densities contain more (extra) information than it is needed to attribute properties to the topological atoms. This may be seen clearly in the case of the atomic charge of a topological atom, as discussed by Popelier and reiterated herein as an example [94]. The total charge density is the sum of the one-electron density and the nuclear point charges, each placed in a separate point in space, $\vec{R}_{\Omega}$, and attributed to a topological atom: $\rho_{total}(\vec{r}) = -\rho(\vec{r}) + \sum_{\Omega} Z_{\Omega} \delta(\vec{r} - \vec{R}_{\Omega})$, where the plus and minus signs are used to distinguish the positive and negative charges, respectively. Integrating the total charge density yields the atomic charge of a topological atom:

$Q_{\Omega} = \int\limits_{\Omega} d\vec{r}\; \rho_{total}(\vec{r}) = -N_{\Omega} + Z_{\Omega}$, where $N_{\Omega}$ is the population of electrons in the atomic basin. Nonetheless, one may use instead the relation between the total charge density and the produced electric field, $\vec{\nabla}\cdot\vec{E}(\vec{r}) = 4\pi\rho_{total}(\vec{r})$, and the Gauss's theorem, $\int\limits_{\Omega} d\vec{r}\left(\vec{\nabla}\cdot\vec{E}\right) = \oint\limits_{\partial\Omega} dS\left(\vec{E}.\vec{n}\right)$, to compute the atomic charge:

$Q_{\Omega} = \left(1/4\pi\right)\oint\limits_{\partial\Omega} dS\; \vec{E}(\vec{r}).\vec{n}(\vec{r})$. The last equation is particularly interesting since it

relies only on the electric field produced on the inter-atomic surfaces of a topological

atom, i.e. the boundaries, to compute the atomic charge, which is usually conceived as a

basin property related to the electronic distribution within the atomic basin. The

subsystem hypervirial theorem for a stationary subsystem is as follows:

$$M_{\Omega} = \mathrm{Re}\left\{(i/\hbar)N\int\limits_{\Omega} d\vec{r}\int d\tau'\; \Psi^*\left[\hat{H},\hat{G}\right]\Psi\right\} = \mathrm{Re}\left\{\oint\limits_{\partial\Omega} dS\left(\vec{J}_G(\vec{r}).\vec{n}(\vec{r})\right)\right\}, \quad \text{where}$$

$\mathrm{Re}$ denotes for the real part, $\hat{G}$ is the one-particle Hermitian *generator* for the property

$\hat{M}$, while $\hat{H}$ and $N$ are the Hamiltonian and the number of electrons of system,

respectively, and $\vec{J}_G(\vec{r})$ is the current of the property $M$ defined as follows:

$$\vec{J}_G(\vec{r}) = \left(N\hbar/2mi\right)\int d\tau'\; \left\{\Psi^*\vec{\nabla}\left(\hat{G}\Psi\right) - \left(\vec{\nabla}\Psi^*\right)\left(\hat{G}\Psi\right)\right\} \qquad (i = \sqrt{-1}) \qquad [1,95].$$

Therefore, the considered example is not something exceptional and the subsystem

hypervirial theorem also entails the same type of the relationship between the "bulk" and

the "boundary" of a topological atom and one may claim that just knowing the currents at

the boundaries is enough to deduce the basin properties. All these are in line with the

idea that the inter-atomic surfaces may be act as *holograms*, i.e. coding the right amount

(not extra) of the information required to deduce basin properties, for the topological atoms. Even more, it has been proposed by Bader and Wiberg that the bond energy between two bonded topological atoms is also derivable from a certain surface integral [96] (see also pages 240-242 in [1]).

While all these seem to be in favor of the idea of the holographic nature of the inter-atomic surfaces, there are properties like the energy of the topological atom, discussed in the previous subsection, which does not seem to be deducible from the known property currents on the inter-atomic surfaces. One may add to this list the electron localization and the delocalization indices of the topological atoms, which are measures of the electron sharing between two atomic basins [97-99] (see also E7.1 appendix in [1]). It seems that the idea of the hologram to be generally true, one needs to derive a general Gauss-like theorem for the topological atoms that goes beyond the subsystem hypervirial theorem and covers all conceivable properties of the topological atoms. The next step will be to answer the question whether such holographic property may be used practically namely, one may derive a theory that only works with the inter-atomic surface, i.e. boundaries/2D surfaces, and yield the properties of the topological atoms without any need to refer to 3D space; this directs us toward the more general idea of *duality* [100-103]. Accordingly, the analytical differential geometry of the inter-atomic surfaces must be also considered more thoroughly [104-108], seeking for general traits of these surfaces that may help to apply the idea of the holography in practice.

As stressed in the introduction, Popelier has given his own interesting insights on the QTAIM some years ago [2], that was inspiring for the present author to embark a research programme on the foundations of the QTAIM. Let us first directly quote these proposals and then discuss their possible implications: "Although no formal comparison has been proposed, it is tempting to think of the atom as a little universe in the sense of the theory of general relativity, loosely speaking. Anyone 'travelling' inside the atom using natural atomic coordinates (see Box 3.1), will have the impression that the atom is infinite. A more careful investigation of the atom's inner curvature, however, should reveal that the atom is bounded in certain regions. Again, differential geometry is the ideal tool to study the intrinsic geometric structure of the atom's space in the hope of linking it to chemical concepts." (see page 43 in [2]) and: "Atoms in molecules are open systems free to exchange charge (and thus matter) and energy with their environment. A challenging programme would be to formulate the thermodynamics of open systems (far from equilibrium) of atom in molecules. This should be possible since the atomic hypervirial theorem defines quantum analogues of classical quantities such as pressure for an atom…" (see pages 107 and 108 in [2]). Both of these proposals are examples of dualities; a topological atom, or more precisely $\rho_\Omega(\vec{r})$, in 3D Euclidian space may "equally" be seen as a free atom, or $\rho_{atom}(\vec{r})$, in a non-Euclidian space [109], and the property currents of a topological atom may "equally" be seen as flows in the boundaries of a stationary but non-equilibrium thermodynamic system [110]. Whether the QTAIM as a theory or the topological atoms as the target objects of this theory may have non-trivial duals as proposed above is an interesting open question that needs further scrutiny.

## 3. Conclusion and prospects

The idea of subsystem is foreign to quantum mechanics and the principles of the theory do not give a clear recipe how a quantum subsystem, or division to system plus environment, must be introduced. Accordingly, this ambiguity opens the door for a large number of quantum subsystem theories, designed for various purposes and applications [89.90]. The fact that these theories do not directly flow from the basic principles of quantum mechanics and contain their own principles and axioms give them their own unique identity, distinct from quantum mechanics; this identity includes their mathematical formalism as well as specific capabilities. Present paper tries to demonstrate that open problems and possible future routes of extension are non-trivial in the case of the QTAIM and unforeseeable developments, like the newly proposed multi-component QTAIM [30-45], are serious possibilities. Interestingly, a detailed reading of Bader's monograph also reveals that he had also proposed some open problems that he believed to be routes for future developments that some, after a quarter of century, have not yet find proper answers. Let us once again quote directly: "Could one develop a liquid model of the charge density, a principle parameter being the viscosity experienced by migrating critical points? As demonstrated in Chapter 6, one can relate the curvature of the charge density at a point in space to the local potential and kinetic energy densities of the electrons. While there is a mechanics of the charge density, it is not complete and the theory of atoms in molecules offers opportunities for the formulation of new models in the search for answers to old problems" (see page 68 in [1] and for another interesting example see pages 218 and 219). All these examples cast doubt that the final chapter of the QTAIM has been written yet and as proposed recently [27,38,79], the QTAIM must be seen as an ongoing "research programme" rather than a "finalized theory".

## Acknowledgments

The author is grateful to Masumeh Gharabaghi and Mohammad Goli for their assistance for preparing the paper.

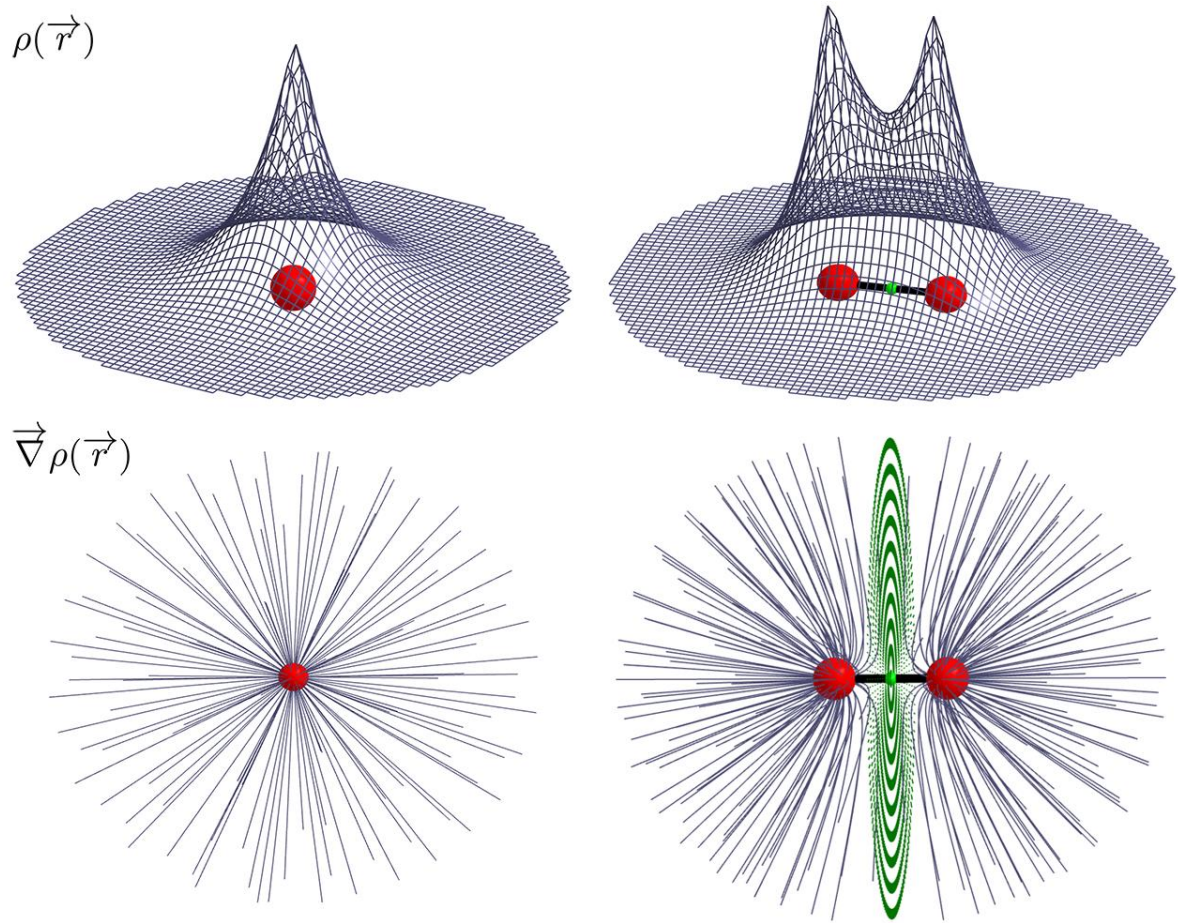

Figure 1- A schematic view of the topological structure of a typical atom and a diatomic molecule composed of two of this atom. The upper panels show the relief maps of the electron density while the large red ball is a schematic representation of the nucleus where (3, -3) attractor is located on whereas (3, -1) critical point is depicted as a small green dot. The lower panels depict the resulting gradient vector field of the electron density where (3, -3) critical points act as the attractor of the field and all the gradient paths in each atomic basin terminate at the attractor. A duo of gradient paths, originating from (3, -1) critical points and terminating at (3, -3) attractors, are depicted as thick black lines. The inter-atomic/zero-flux surface is the green flat surface perpendicular to the black lines and going through (3, -1) critical point. All the gradient paths on this surface terminate at (3, -1) critical point.